# Engineering Cooperative JADE Agents with the AMCIS Methodology: The Transportation Management Case Study

Djamel Benmerzoug, Salim Djaaboub, and Hani Mahmoudi

Department of Computer Science,
Mentouri University of Constantine 25000, Algeria
Tel.: + 213 31 63 90 10
E-mail : {benmerzougdj, dja_salim}@yahoo.fr

**Abstract.** This paper discusses in detail important analysis and design issues emerged during the development of an agent-based transportation e-market. This discussion is based on concepts coming from the AMCIS methodology and the JADE framework. The AMCIS methodology is specifically tailored to the analysis and design of cooperative information agent-based systems, while it supports both the levels of the individual agent structure and the agent society in the Multi-Agents Systems (MAS) development process. According to AMCIS, MAS are viewed as being composed of a number of autonomous cooperative agents that live in an organized society, in which each agent plays one or more specific roles, while their plans and interaction protocols are well defined. On the other hand JADE is a FIPA specifications compliant agent development environment that gives several facilities for an easy and fast implementation. Our aim is to reveal the mapping that may exists between the basic concepts proposed by AMCIS for agents specification and agents interactions and those provided by JADE for agents implementation, and therefore to propose a kind of roadmap for agents developers.

## 1 Introduction

In the last few years, multi-agent systems (MAS) have successfully addressed a variety of business problems, such as business process management [10], supply chain management [6], and manufacturing scheduling [8]. The above success is partially due to the nature of software agents, which autonomously perform well-defined activities, based on rules and procedures coded into their behaviour. It is widely argued that such systems become more powerful and may solve complex problems when groups of different kinds of agents begin communicating in an efficient and effective way [9]. Indeed, several efforts have been made to develop agent-based software methodologies [3, 11]. However, most of the studies concentrate on specific areas of the agent software – agent models, reasoning logic and agent actions, agent communication, agent programming languages and frameworks. To gain wide acceptance of agent-based software in practice, methods covering the full software development cycle from analysis to implementation are required. To this end, we have previously developed an agent-oriented method named AMCIS (an **A**gent-Oriented **M**ethod for Designing

Cooperative Information Systems) for the development of applications that support inter-agent cooperation [3, 4].

In this paper we present an attempt to use the AMCIS method in order to engineer a MAS that is to be implemented with the JADE framework. The only pretension we have with this paper is to share our experience to conceive and develop a MAS, by combining AMCIS and JADE, in the context of transportation management.

Some architectures of transportation management e-market have been provided and many technologies have been applied in this field [7]. however, despite their efficient management of the underlying databases (where information about transport companies, costs, itineraries etc. is kept), the major drawback of these approaches is the lack of the methodological aspect for the design of this kind of applications. It is then frequent to certify that the developed systems put numerous problems. They don't always satisfy the users' needs, and don't provide any tools and methods to identify the requirements of the system to be. So how to analyze the requirement of e-market application is the foundation stone for modelling it.

This paper discusses in detail important analysis and design issues emerged during the development of our agent-based transportation e-market. This discussion is based on concepts coming from the AMCIS methodology and the JADE framework. To briefly introduce these approaches here, we only mention that the AMCIS methodology is specifically tailored to the analysis and design of cooperative information agent-based systems, while it supports both the levels of the individual agent structure and the agent society in the MAS development process. According to AMCIS, MAS are viewed as being composed of a number of autonomous cooperative agents that live in an organized society, in which each agent plays one or more specific roles, while their plans and interaction protocols are well defined (for more details, see [3, 4]).

On the other hand, JADE is a software development framework [2], fully implemented in Java, which aims at the development of multi-agent systems and applications that comply with FIPA standards (http://www.fipa.org). JADE provides standard agent technologies and offers to the developers a number of features in order to simplify the development process. Following JADE, agent tasks and intentions are implemented through the use of behaviours (for more details on this issue, see http://sharon.cselt.it/projects/jade).

The proposed Ag-Market (AGent-based transportation e-MARKET) system able to efficiently handle transportation transactions of various types. Agents of our system represent and act for any user involved in a transportation scenario, such as customers who look for efficient ways to ship their products and transport companies that may - fully or partially - carry out such requests, while they cooperate and get the related information in real-time mode. Our overall system is based on software agents that achieve efficient communication among all parties involved, coordinate the overall process, construct possible alternative solutions and perform the required decision-making.

The remainder of this paper is organized as follows. Section 2 presents a quick overview of the AMCIS method. Section 3 discusses in detail important analysis and design issues emerged during the development of the Ag-Market system, while section 4 presents some implementation aspects. Finally, concluding remarks and future work directions are given in section 5.

## 2 AMCIS Overview

In the context of cooperative applications (particularly e-commerce) we belief that the main challenge will shift towards the understanding of business frame and needs, and how to make decisions involving technical systems to address those needs and concerns. To this end, we need a clearer understanding of what it means for systems to be "cooperative". So, the notion of cooperation cannot be fully addressed unless the goals and desires of agents are explicitly modelled and argued about. According to these perspectives, we have developed the AMCIS method which represents a methodological agent-based framework in order to guide the design of cooperative systems. It aims to analyse then to model the cooperative processes among a set of agent in measure to answer their needs.

AMCIS identifies three phases of design, which operate in two levels of abstraction (Figure 1). The *Analysis* phase produces an understanding about organizational relationships and the rationales behind them. The *System Design* and *Agent Design* phases focus on the system specification, according to the requirements resulting from the above phase.

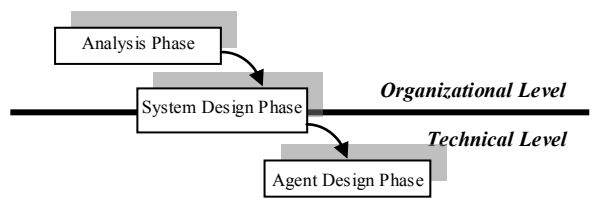

**Fig. 1.** Different phases of the AMCIS method

- ***Organizational level***: this level corresponds to the modelling of the cooperation during the description of business frame. That is, the modelling of the cooperation among the various organizations in term of a set of actors, which depend on each other for goals to be achieved, tasks to be performed, and resources to be furnished. By clearly defining these dependencies, it is then possible to state the why, beside how and which are the actors entering in an operation requiring their cooperation. This level defines the system's global architecture in terms of subsystems, interconnected through data and control flows.
- ***Technical level:*** this level deals with the specification of the agents' micro level. Then, capabilities, plans, and the communication among agents are specified in detail.

The following section presents a quick overview of the proposed Ag-Market system and its architecture and discuss in detail important analysis and design issues emerged during the development of this system.

## 3   Using AMCIS for Analysis and Design of the Ag-Market

The Ag-Market is a distributed agent-oriented implementation of transportation management e-market, where agents work in an open environment as Internet (or Intranet). The proposed system (see Fig. 2) will offer efficient mechanisms for the matching of requests and offers for transport services, as well as for the construction of feasible solutions. As it will be discussed in more detail in the sequel, the Ag-Market system is equipped with the aid of three types of agents (namely, the *Broker*, the *Customer* and the *Provider* agents). After a matching has been found, the system will dynamically inform the related users with the alternative possibilities found to cover their requests. In this case, our system considered as a multi-criteria decision support tool, through which buyers may evaluate alternative solutions. More precisely, a candidate customer will be able to evaluate, through a set of criteria (delivery time, cost, insurance etc.), the alternative offers found, to express preferences and constraints concerning these criteria, and to weigh them accordingly. Moreover, customers may amend some of the features of the offers received, the aim being to further negotiate on them with the carriers. Through continuous interaction with the users, the system will propose the best solution according to the existing information.

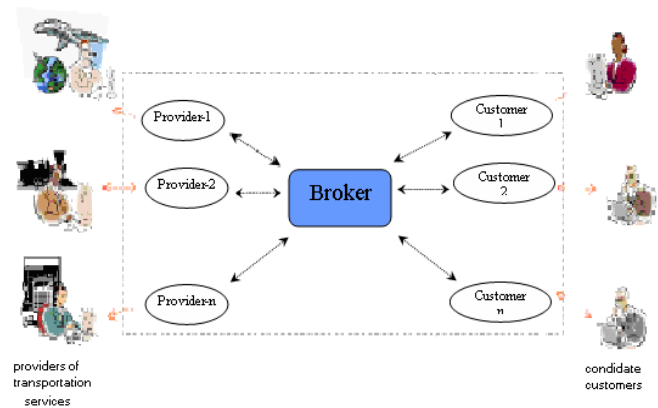

**Fig. 2.**  The proposed Ag-Market

The following section discusses in detail important analysis and design issues emerged during the development of our agent-based transportation e-market.

### 3.1 Analysis Phase

Following the AMCIS methodology, the basic objective in the analysis of our system was the understanding of organizational relationships and the rationales behind them. This phase defines the system's global architecture in terms of subsystems, interconnected through data and control flows. Indeed, AMCIS proposes to use the notion of strategic actor based on the i* framework [12] to represent the cooperation in the Ag-Market system. This system is viewed as consisting of social actors who have freedom of action, but depend on each other for goals to be achieved, tasks to be performed, and resources to be furnished. The i* framework is very useful to provide a higher level modelling method comparing to conventional modelling techniques such as data flow diagramming and object-oriented analysis. It includes the strategic dependency (SD) model for describing the network of relationships among actors, as well as the strategic rationale (SR) model for describing and supporting the reasoning that each actor has about its relationships with other actors [12].

#### 3.1.1   Construction of the SD model
The SD model attempts to define the cooperation between the entities of the system to attain global objective. The Figure 3 shows the SD model for the example of our case. The SD model is a graph where each node represents an *actor*, and each link between two actors indicates that one actor depends on the other for something in order that the former may attain some goal. In [12], the depending actor is called the *depender*, and the actor who is depended upon is called the *dependee*. The object around which the dependency relationship centers is called *dependum*. Authors distinguish among four types of dependencies, based on the type of the dependum: *hardgoal dependency*, *softgoal dependency, task dependency* and *resource dependency*. The softgoal having no clear-cut definition and/or criteria for deciding whether they are satisfied or not. Softgoal are typically used to model non-functional requirements. For simplicity, in the rest of the paper goal refer to hardgoal when there is no danger of confusion.

In figure 3, the customer depends on the Broker to obtain a set of solution. In the other hand, the Broker depends on the Customer to obtain the solution (itinerary) selected. It can also depend on the Provider to confirm the itinerary selected by the Customer.

#### 3.1.2   Construction of the SR model
The SR model describes a more detailed level of an actor in order to model its internal intentional relationship and support to reason of each actor about its intentional relationships. The intentional elements, such as goals, tasks, resources and softgoals, appear in SR model not only as external dependencies, but also as internal elements arranged into a hierarchy of means-ends and task-decompositions relationships. A goal may be associated through means-ends with multiple, alternative ways to achieve it, which are usually represented as tasks. Task decomposition links hierarchically decompose task into four types: sub-tasks, subgoals, resources, and softgoals [12]. In fi-gure 4, the goals and internal intentional relationships of the Ag-Market are elaborated.

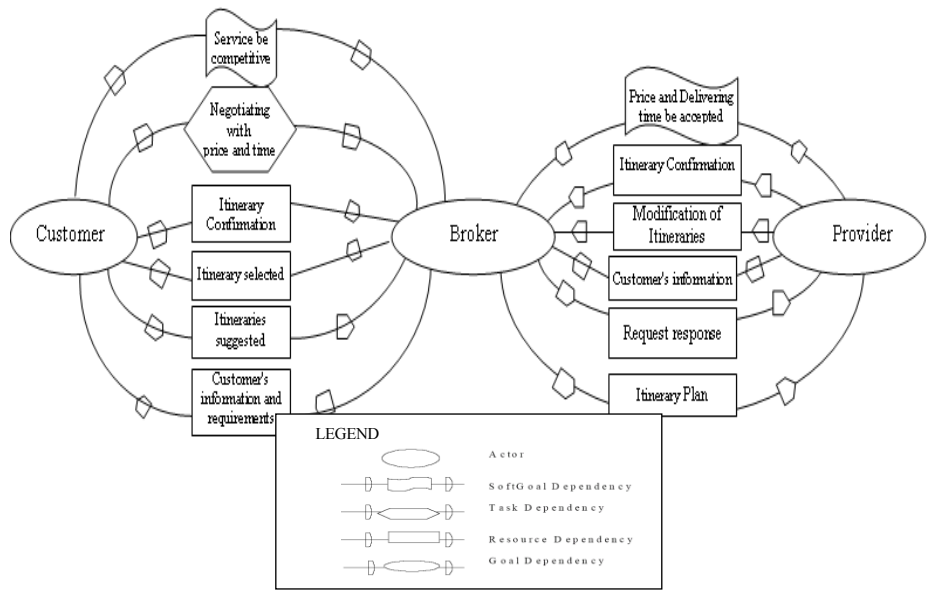

**Fig. 3.** The SD model

The goal of the Customer is to get an itinerary and is accomplished by "`Itinerary be decided`". This task can be decomposed of two sub-tasks, such as "`Requirements of service be decided`" and "`Select a proposition`". For the Broker agent, it depends on the Provider agent in order to define route paths that successfully address a customer's request. In the other hand, the Provider agent depends on the Broker agent to get the customer's requirements and for updating its Itinerary plan. This agent has tow main tasks: "`Treatment of reservation request`" and "`Construction of the Itinerary plan`". For the second task, there are two ways possible to carry out the demand of the Broker (first by "`Plan Construction`" and second by "`Plan Updating`").

### 3.2  System Design Phase
In this phase, the developer begins the design of its system. He starts with a global design without entering in details. This phase consists of identifying capacities of each agent, and the links among them (acquaintances model).

#### 3.2.1 Identification of capacities of each agent
This step consists of the identification of the capacities needed by the agents to fulfill their goals and satisfy their dependencies. Following the approach described in [5], capacities can be easily identified by analysing the SR model. In particular, each dependency relationship can give place to one or more capacity trigged by external events. To give an intuitive idea of this process let's focus on a specific actor of the SR model, such as the Broker, and we consider all the in-going and out-sorting dependencies, as shown in figure 4. Each dependency is mapped to a capacity. So, for instance, the dependency for the resource "`Customer Requirements`" calls the capacity "`Get Customer Requirements`", and so on.

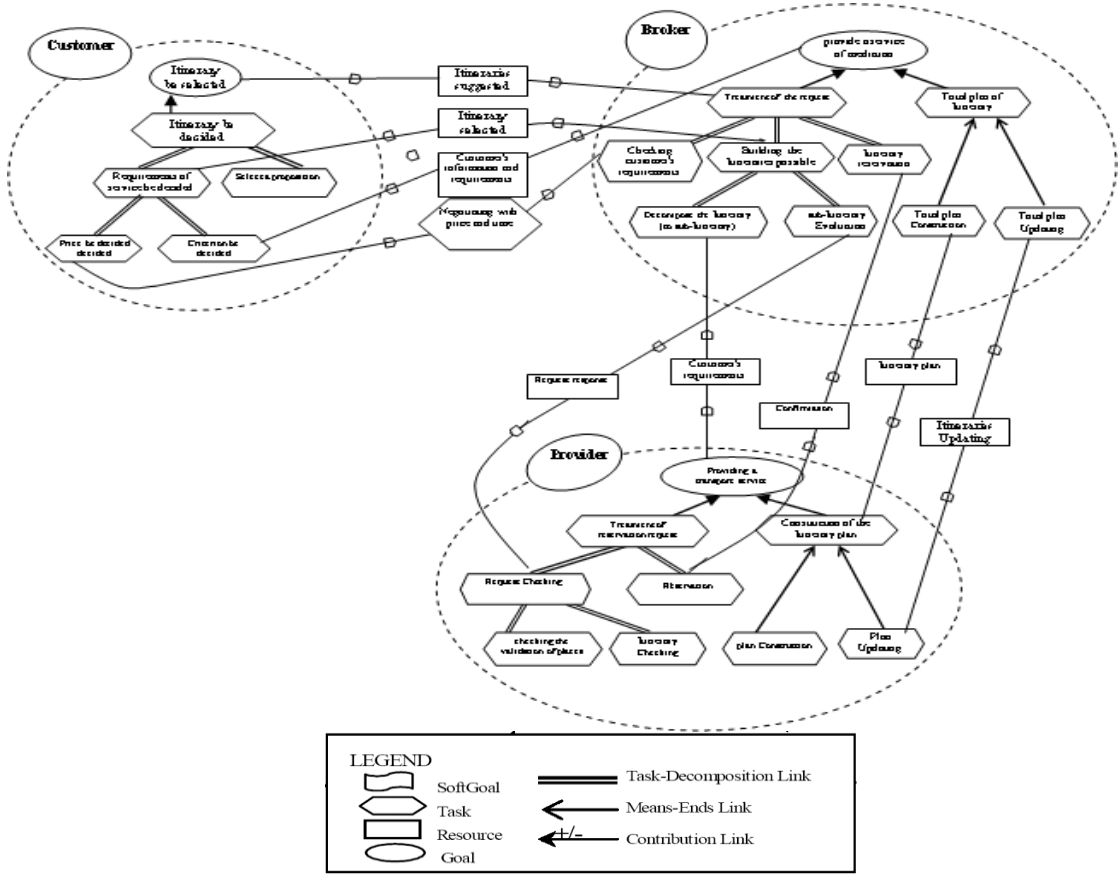

**Fig. 4.** The SR model

### 3.2.2 Construction of the acquaintances model
According to [11], an agent acquaintance model (figure 5) is simply a graph, with the nodes corresponding to agent types loosened in the first stage of this phase, and arcs in the graph corresponding to communication pathways. Agent acquaintance models are directed graphs, and so an arc a→b indicates that a will send messages to b, but not necessarily that b will send messages to a. An acquaintance model may be derived in a straightforward way from the SR model.

### 3.3 Agent Design Phase
This phase allows better to detail initial design and to refine the system design. It is constituted of two steps, each of them taking charge of the construction of a model; the interactions model and the plans model. During this phase, we adopt a subset of the AUML (for Agent Unified Modelling Language) sequence diagrams proposed in [1] for specifying interactions among agents (interaction model) and the activity diagrams for representing agents plans (plans model).

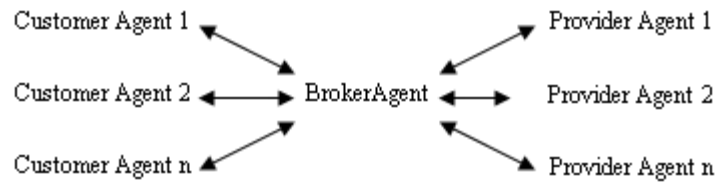

**Fig.5.** The acquaintance model

### 3.3.1 Construction of the Interaction Model

In AUML sequence diagrams, agents correspond to objects, whose life-line is independent from the specific interaction to be modelled (in UML an object can be created or destroyed during the interaction); communication acts between agents correspond to asynchronous message arcs [1].

Figure 6 shows a simple part of the interaction model. In particular, the diagram models the interaction among the Costumer agent, the Broker agent and the Provider agent. The interaction starts with a request by the Customer, and ends with the results presentation by the Broker agent to the Customer.

### 3.3.2 Construction of the Plans Model

The plans model consists of a set of plans, each being specified with an activity diagram. This one expresses how an agent has to behave to react to an event. The figure 7 presents the `Evaluate_Customer_Requirements` diagram plan of the Broker agent when it receives an event (*inform*) from the Customer agent. In this diagram ovals represent capabilities and arcs represent internal or external events.

In the following section of this paper, we describe the mapping that may exist between the basic concepts proposed by our method for agents specification and agents interactions and those provided by JADE for agents implementation.

## 4  Some Implementation Aspects

The platform chosen for the implementation is Java Agent Development framework (JADE) [2]. JADE is a software development framework, fully implemented in Java, which aims at the development of multi-agent systems and applications that comply with FIPA standard. To achieve such a goal, JADE offers the following list of features to the agent programmer:
− Distributed agent platform. The agent platform can be distributed on several hosts, each one of them executes one Java Virtual Machine.

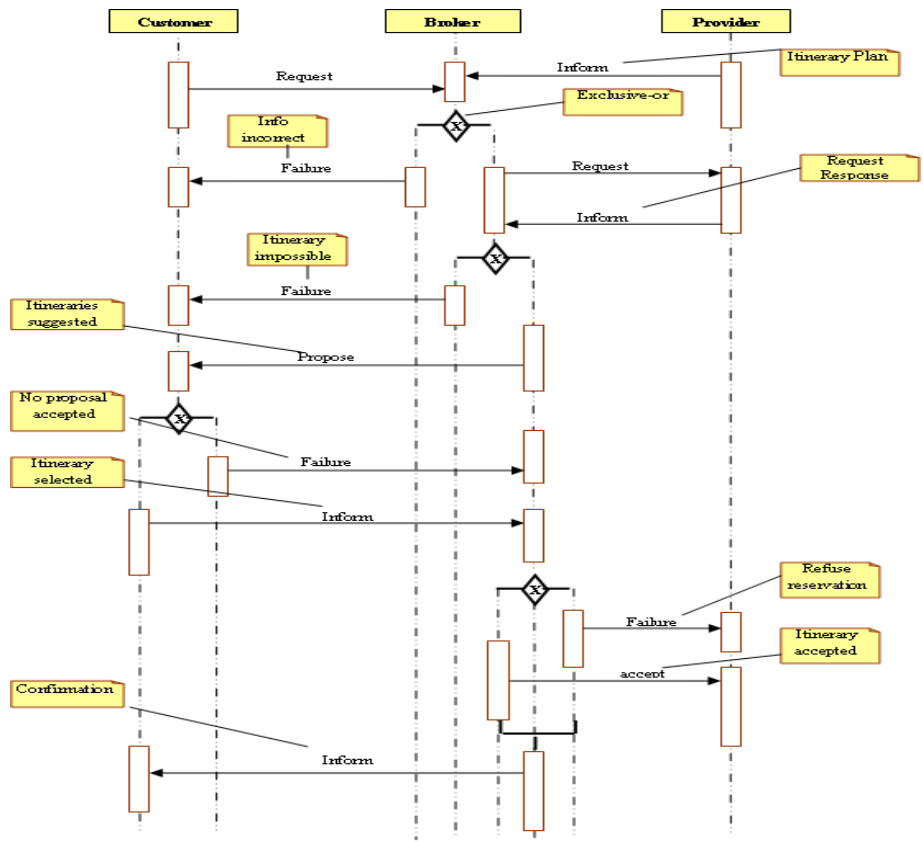

**Fig.6.** The AUML interaction diagram

- FIPA-Compliant agent platform, which includes the Agent Management System, the Directory Facilitator and the Agent Communication Channel.
- Efficient transport of ACL messages between agents.

### 4.1 Mapping AMCIS Concepts into JADE Code

Creating a JADE agent is as simple as defining a class extending the *Jade.core.Agent* class and implementing the *setup()* method. The *setup()* method is intended to include agent initializations. The actual job an agent has to do is presented as JADE behaviours. Besides, the predefined behaviour `PerceptionInterface` is started by *setup()* as shown in the JADE code below.

```
import jade.core.*;
public class MyAgent extends Agent {
 // Possible variables
 protected void setup {
   // to do: add necessary start-up code
      addBehaviour(new PerceptionInterface(this)); }
  }
```

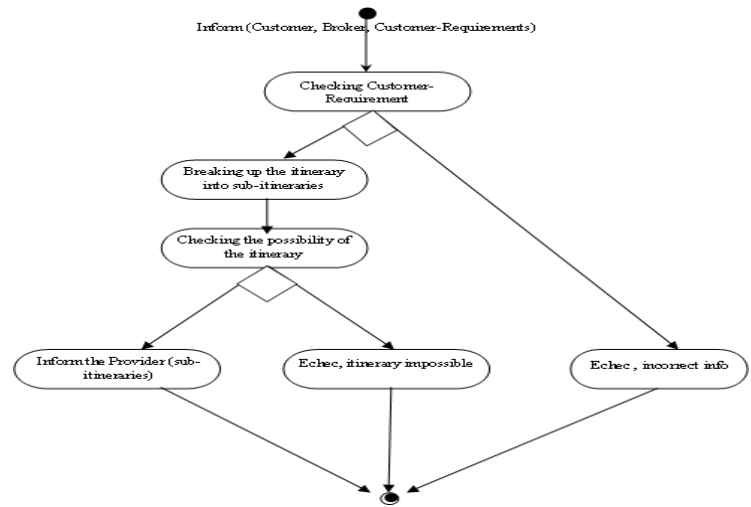

**Fig. 7.** The Evaluate_Customer_Requirements Plan

In our case, we propose to implement capabilities resulting from system design phase as *SimpleBehaviour* and plan resulting from agent design phase as *FSMBehaviour* class. The *FSMBehaviour* class is a *CompositeBehaviour* that executes its children (sub behaviour that are going to be used by *FSMBehaviours*) according to a Finite State Machine defined by the user.

The perception interface is derived from JADE class *CyclicBehaviour* and will therefore run continuously. If no messages have arrived, the behaviour will block and restart after a new message has arrived. If a message has arrived, the perception interface has to interpret this message into a set of goals, starts `PlanRetrival` behaviour and resumes waiting for incoming messages. The `PerceptionInterface` behaviour code is shown below.

```
import java.util.*;
import jade.core.*;
import jade.core.behaviour.*;
import jade.lang.acl.ACLMessage.*;
class PeceptionInterface extends CyclicBehaviour {
  public PeceptionInterface (Agent a) {super(a);}
  Vector goals = new Vector();
  public void action() {
    // wait for message
    ACLMessage received = myAgent.receive();
    if(ACLMessage == null) {block();}
    else {
      // message interpretation
      ………………………
      // Start PlanRetrieval Behaviour
      addBehaviour(new PlanRetrival(this,goals));
        }
      }
   }
```

The `PlanRetrival` behaviour is derived from JADE class *SimpleBehaviour*. This behaviour takes as parameters a set of goals (the results of message interpretation by perception interface) and finds the adequate plan from the plans library.

### 4.2 Inter-Agent Communication

Adopting JADE, communication between agents is facilitated through predefined methods. For instance, agents may identify themselves with a unique name (*setName(AID)* method of *DFAgentDescription* class), get registered to the JADE platform (*register(this, DFAgentDescription)* method of *DFService* class), and communicate using the *send(message)* and *receive(message)* methods of the *ACLMessage* class. JADE offers the possibility to follow the communication between agents by means of the "sniffer" agent which is a GUI whose output is shown in figure 8 (the communication between Customer agent, Broker Agent and tow Provider agents).

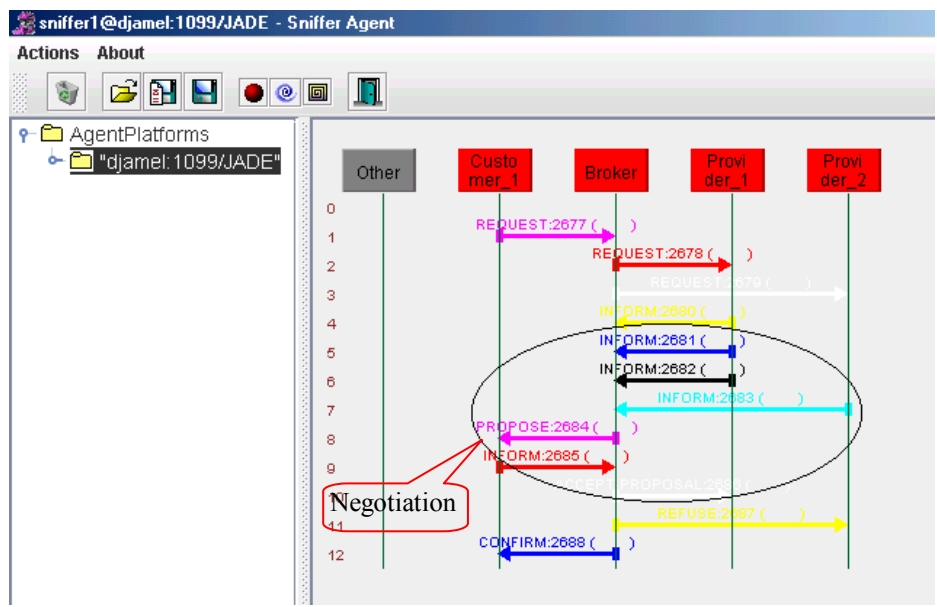

**Fig. 8.** Output of the JADE Sniffer agent

## 5 Conclusion and Future Work

In this paper we have presented the analysis, design and implementation phases of an agent-based transportation e-market. As we already have said before, the only pretension we have with this paper is to share our experience on how one can combine the AMCIS methodology and the JADE development environment in order to implement a real multi-agent system. AMCIS methodology is an easy to use agent-orient software development methodology that however presently, covers only the phases of analysis and design. On the other hand JADE is a FIPA specifications compliant agent development environment that gives several facilities for an easy and fast implementation. Our aim was to reveal the mapping that may exists between the basic concepts proposed by AMCIS for agents specification and agents interactions and those provided by JADE for agents implementation, and therefore to propose a kind of roadmap for agents developers.

Our primary future work direction is certainly the *full* implementation of the electronic marketplace presented in this paper. Another direction concerns research on alternative ways of coordination of customers and providers in our framework. Finally, further work needs to be done on the definition of semantics and syntax of data and knowledge conveyed among our agents.

## 6 References

1. Bauer, B., Müller, J., Odell, J.: Agent UML: A Formalism for Specifying Multiagent Interaction. Agent-Oriented Software Engineering, Springer, Berlin, (2001) 91-103
2. Bellifemine, F., Caire, G., Trucco, T., Rimassa, G.: Jade Programmer's Guide. JADE 3.0b1. Available at http://sharon.cselt.it/projects/jade/, (2003)
3. Benmerzoug, D., Boufaida, Z., Boufaida, M.: From the Analysis of Cooperation Within Organizational Environments to the Design of Cooperative Information Systems: An Agent-Based Approach. R. Meersman et al. (Eds.): OTM Workshops 2004, LNCS Springer-Verlag, (2004) 496-506
4. Benmerzoug, D., Boufaida, M., Boufaida, Z.: Developing Cooperative Information Agent-Based Systems with the AMCIS Methodology. IEEE International Conference on Advances in Intelligent Systems: Theories and Application. Luxembourg, (2004)
5. Bresciani, P., Giorgini, P., Giunchiglia, F., Mylopoulos, J., Perini, A.: TROPOS: An Agent-Oriented Software Development Methodology. Accepted in JAAMAS, (2003)
6. Fox, M.S., Barbuceanu, M., and Teigen, R.: Agent-Oriented Supply-Chain Management. International Journal of Flexible Manufacturing Systems 12(2/3), (2000) 165-188.
7. Nikos Karacapilidis, Alexis Lazanas, Pavlos Moraïtis.: an agent-mediated marketplace for transportation transactions. Proc in ICEIS'03, (2003) 238-243
8. Shen W. and Norrie, D.H.: An Agent-Based Approach for Dynamic Manufacturing Scheduling. In Proceedings of Autonomous Agents'98 Workshop on Agent-Based Manufacturing, Minneapolis/St. Paul, MN, (1998) 117-128.
9. Sycara, K., and Zeng, D., 1996. Coordination of Multiple Intelligent Software Agents. International Journal of Cooperative Information Systems 5(2-3).
10. Jennings, N., Faratin, P. Norman, T.J., O'Brien, P. and Odgers, B., 2000. Autonomous Agents for Business Process Management. International Journal of Applied Artificial Intelligence 14(2), 145-189.
11. Wooldridge, M., Jennings, N., Kinny, D.: The GAIA Methodology for Agent-Oriented Analysis and Design. Journal of Autonomous Agents and MAS, (2000) 285-312
12. Yu, E.: Agent Orientation as a Modelling Paradigm. Wirtschaftsinformatic. 43(2), (2001) 123-132